\documentstyle[preprint,graphicx]{jpsj}
\newcommand{\eps}{\varepsilon}

\newcommand{\vct}[1]{\mbox{\boldmath #1}}
\def\runtitle{Origin of G-type Antiferromagnetism 
and Orbital-Spin Structures in ${\rm LaTiO}_3$}
\def\runauthor{Masahito {\sc Mochizuki} and Masatoshi {\sc Imada}}
\title{Origin of G-type Antiferromagnetism 
and Orbital-Spin Structures in ${\rm LaTiO}_3$  
}
\author
{Masahito {\sc Mochizuki} and Masatoshi {\sc Imada}}
\inst
{Institute for Solid State Physics, 
University of Tokyo,\\ 5-1-5 Kashiwa-no-ha, Kashiwa, Chiba 277-8581}
\recdate{\today}

\abst{ The possibility of the $D_{3d}$ distortion of
${\rm TiO}_6$ octahedra is examined 
theoretically in order to understand the origin 
of the G-type antiferromagnetism (AFM(G))
and experimentally observed puzzling 
properties of ${\rm LaTiO}_3$.
By utilizing an effective spin and pseudospin Hamiltonian
with the strong Coulomb repulsion, it is shown
that AFM(G) state is stabilized through the 
lift of the $t_{2g}$-orbital degeneracy
accompanied by a tiny $D_{3d}$-distortion .
The estimated spin-exchange interaction is in agreement 
with that obtained by the neutron scattering.
Moreover, the level-splitting energy due to the distortion 
can be considerably larger than the spin-orbit interaction
even when the distortion becomes smaller than the detectable limit 
under the available experimental resolution. 
This suggests that the orbital momentum 
is fully quenched and the relativistic
spin-orbit interaction is not effective in this system, 
in agreement with recent neutron-scattering experiment.}

\kword
{${\rm LaTiO}_3$, ${\rm GdFeO}_3$-type distortion, 
$D_{3d}$-crystal field, orbital ordering,
second-order perturbation theory, G-type antiferromagnetism}
\begin{document}
\sloppy
\maketitle
%
% introduction
%
\section{Introduction} 

\begin{figure}[tdp]
%\hfil
\includegraphics[scale=0.25]{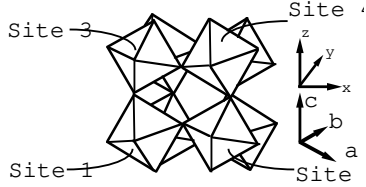}
%\epsfile{file=gdfo3.eps,scale=0.2}
%\hfil
\caption{${\rm GdFeO}_3$-type distortion.}
\label{gdfo3}
\end{figure}

Perovskite-type Ti oxide $R{\rm TiO}_3$ 
have recently attracted
much interest because of the rich magnetic and orbital phases
caused by an interplay of spin and orbital degrees of freedom.~\cite{Imada98} 
In these compounds, 
${\rm Ti}^{3+}$ has a $3d^1$ configuration, and one of the threefold 
$t_{2g}$-orbitals is occupied at each transition-metal 
site. The crystal structure is a pseudocubic perovskite with
${\rm GdFeO}_3$-type distortion
in which the ${\rm TiO}_6$ octahedra forming the perovskite lattice
tilt alternatingly.
The unit cell contains four ${\rm TiO}_6$-octahedra, 
as shown in Fig.~\ref{gdfo3}.
%The magnitude of the distortion depends on the ionic radii
%of the $R$ ion. 
With a smaller ionic radius of the $R$ ion, 
the distortion increases with the decrease in the Ti-O-Ti bond angle
from $180^{\circ}$.
The bond angle can be controlled by use of the solid-solution systems  
${\rm La}_{1-y}{\rm Y}_{y}{\rm TiO}_3$
or in $R{\rm TiO}_3$, by varying the $R$ ions.
%In particular, with varying the Y concentration in 
%${\rm La}_{1-y}{\rm Y}_{y}{\rm TiO}_3$, we can control the 
%bond angle almost continuously from $157^{\circ}$
%($y = 0$) to $140^{\circ}$ ($y = 1$).
%~\cite{MacLean79}

Recently, the magnetic phase diagrams have been investigated
intensively in the plane of temperature and the magnitude of the
distortion.~\cite{Okimoto95,Katsufuji97,Goral82,Greedan85}
In particular, ${\rm LaTiO}_3$ exhibits a G-type AFM (AFM(G)) ground state
with magnetic moment of 0.45 $\mu_{\rm B}$,~\cite{Goral82} and the
N${\rm {\grave{e}}el}$ temperature ($T_N$) takes about 130 K.
With increasing the ${\rm GdFeO}_3$-type distortion,
$T_N$ decreases rapidly and is suppressed to almost zero,
subsequently a ferromagnetic (FM) ordering appears.
In Y-rich systems and in ${\rm YTiO}_3$ in which the 
${\rm GdFeO}_3$-type distortion is relatively large, the system shows
a FM ground state accompanied by a large Jahn-Teller (JT) distortion.
Recently, the AFM-FM phase transition with 
the second-order-like behavior has been understood from 
the strong two-dimensionality in the spin-coupling realized near
the transition point.~\cite{Mochizuki00,Mochizuki01a}
However, the origin and nature of AFM(G) state in the 
end compounds ${\rm LaTiO}_3$ remains unresolved.

Recent model Hartree-Fock study predicts
the AFM(G) state with the spin-orbit ground state
%out of which two states $\frac{1}{\sqrt{2}}(yz+izx)\uparrow$
%and $\frac{1}{\sqrt{2}}(yz-izx)\downarrow$ 
%are alternating between nearest neighbors
in the absence of static 
JT distortion.~\cite{Mizokawa96a,Mizokawa96b}
However, in contrast to the prediction
from the spin-orbit ground state, recent neutron-scattering study shows 
the spin-wave spectrum well described by an
isotropic spin-1/2 Heisenberg model 
with a nearest-neighbor superexchange constant 
$J\sim15.5$ meV.~\cite{Keimer00}
This also suggests the absence of unquenched orbital moments.
When the orbital degeneracy remains, a FM spin structure
with antiferro-orbital ordering in which 
the neighboring orbitals are orthogonal to each
other is eventually expected in the presence of the transfers and the 
Coulomb-exchange interaction.
Indeed, recent weak coupling study shows
that in the cubic-crystal field in the absence of the static JT distortion, 
a FM state, out of which 
two states $\frac{1}{\sqrt{2}}(yz+izx)\uparrow$ and $xy\uparrow$
are alternating is favored both by the spin-orbit interaction 
and by the spin-orbital superexchange interaction.~\cite{Mochizuki01b}
However, no evidence for the orbital ordering is found 
in a resonant x-ray scattering study.~\cite{Keimer00}

Recently, a possible orbital liquid state was proposed
on the basis of small exchange interaction in the orbital
sector in the AFM(G) state.~\cite{Khaliullin00}
However, in a system with strong orbital 
fluctuations, the spin and orbital degrees
of freedom strongly couple, and both degrees of freedom
can not be determined independently to each other.
In this circumstance, the origin of AFM(G) state in ${\rm LaTiO}_3$
is to be clarified in a self-consistent manner.

\begin{figure}[tdp]
\hfil
\includegraphics[scale=0.35]{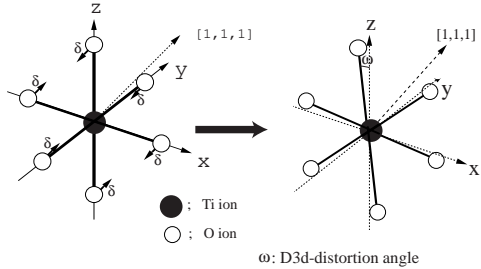}
\hfil
\caption{$D_{3d}$ distortion.}
\label{D3ddist}
\end{figure}

In this letter, in order to understand the above
properties of AFM(G) state in ${\rm LaTiO}_3$,
we propose that the puzzle is solved if we
assume the existence of a small $D_{3d}$-crystal field.
If the $D_{3d}$ distortion exists, ${\rm TiO}_6$-octahedron is contracted
along the threefold direction as shown in Fig.~\ref{D3ddist}.
As a result, threefold degenerate $t_{2g}$-levels split
into non-degenerate lower $a_{1g}$-level ($\varphi$1) 
and twofold-degenerate higher $e_g$-levels ($\varphi$2, $\varphi$3).
In the distortion with [1,1,1]-trigonal axis,
the $a_{1g}$ representation is $\frac{1}{\sqrt{3}}(xy+yz+zx)$
and the $e_g$ representations are 
$\frac{1}{\sqrt{6}}(2xy-yz-zx)$ and
$\frac{1}{\sqrt{2}}(yz-zx)$.
Since the lowest level has no degeneracy, this distortion 
significantly and sensitively lowers the energy of the electron system
in $t_{2g}^1$ configuration.
Though the $d$-type JT distortion is stabilized in ${\rm YTiO}_3$ 
or in the compounds with large ${\rm GdFeO}_3$-type distortion
due to the effects of covalency between
O $2p$ and unoccupied $d$-orbitals on the $R$-site cations,~\cite{Mizokawa99}
${\rm LaTiO}_3$ with small ${\rm GdFeO}_3$-type distortion
turns out to be unstable to the $D_{3d}$ distortion.   
The magnitude of the  distortion can be denoted by the $D_{3d}$-distortion
angle $\omega$ (see Fig.~\ref{D3ddist}).
By utilizing the effective spin and pseudospin Hamiltonian
in the strong Coulomb repulsion, we solve the self-consistent
Hartree-Fock equation and show that the 
emergence of the AFM(G) state and the following features in ${\rm LaTiO}_3$
are well explained by the effects of $D_{3d}$ distortion
even when $\omega$ is as small as $1^{\circ}$:
\begin{itemize}
\item $T_N \sim$130 K, $J \sim 15.5$ meV,
\item magnetic moment of 0.45 $\mu_{\rm B}$,
\item no detectable JT distortion,
\item no evidence for orbital ordering in the resonant x-ray scattering.
\end{itemize}
We analyze why the orbital ordering has not been detected so far, 
and propose experimental ways for testing the validity of our proposal. 
\section{Formalism}
 
We start with the multiband $d$-$p$ model in which 
the full degeneracies of Ti $3d$ and O $2p$
orbitals as well as the on-site Coulomb and exchange 
interactions are taken into account.~\cite{Mochizuki01a}
The nearest-neighbor $d$-$p$ and $p$-$p$ transfers are  given in terms of 
Slater-Koster parameters $V_{pd{\pi}}$, $V_{pd{\sigma}}$, $V_{pp{\pi}}$ 
and $V_{pp{\sigma}}$.~\cite{Slater54} 
The on-site Coulomb interactions are expressed using Kanamori parameters,
$u$, $u^{\prime}$, $j$ and $j^{\prime}$ which 
satisfy the following relations:~\cite{Brandow77,Kanamori63}
$u = U + \frac{20}{9}j$, $u'= u -2j$ and $j = j'$.
Here, $U$ gives the magnitude of the multiplet-averaged 
$d$-$d$ Coulomb interaction.
The charge-transfer energy $\Delta$, which describes the energy
difference between occupied O $2p$ and unoccupied
Ti $3d$ levels, is
defined by using $U$ and energies of the
bare Ti $3d$ and O $2p$ orbitals $\eps_d^0$ and $\eps_p$
as $\Delta = \eps_{d}^0 + U -\eps_p$,
since the characteristic unoccupied $3d$ level energy on the
singly occupied Ti site is $\eps_{d}^0 + U$.
The values of the parameters are
estimated by the cluster-model analyses of valence-band and
transition-metal $2p$ core-level photoemission spectra
and the analyses of the first-principle band
calculations.~\cite{Saitoh95,Mahadevan96}
We take the values of these parameters as
$\Delta = 5.5$ eV, $U = 4.0$ eV, $V_{pd\sigma} = -2.4$ eV,
$V_{pd\pi} = 1.3$ eV, $V_{pp\sigma} = 0.52$ eV, 
$V_{pp\pi} = -0.11$ eV and $j = 0.46$ eV
throughout the present calculation. 
The effects of the ${\rm GdFeO}_3$-type distortion 
are considered through the 
transfer integrals which is defined by using the
Slater-Koster's parameters~\cite{Slater54}.
In the present calculations, we simulate the ${\rm GdFeO}_3$-type distortion 
by rotating the ${\rm TiO}_6$ octahedra by angle $+14.1^{\circ}$ and
$-14.1^{\circ}$ around the [1,1,1] and [-1,-1,1] axes.
%with respect to the $x$, $y$, and $z$ axes.
As a result, the Ti-O-Ti bond angle takes $\sim 157^{\circ}$.
The effects of the $D_{3d}$ distortion are also considered. 
We integrate out the O $2p$ orbital degrees of freedom 
in the path-integral formalism to arrive at the effective 
multiband Hubbard model which includes only Ti $3d$ orbital degrees of freedom.
 Starting with thus obtained multiband Hubbard Hamiltonian, 
we can derive an effective Hamiltonian in the low-energy region
on the subspace of states
with singly occupied $t_{2g}$ orbitals
at each transition-metal site by utilizing a 
second-order perturbation theory.
The states of $3d$ electrons localized at the transition-metal
sites can be represented by two quantum numbers, the 
$z$-component of the spin $S_z$
and the type of the occupied orbitals. 
The threefold degeneracy of $t_{2g}$-orbitals
at each site can be described by a pseudospin 
with the spin-1 operator $\vct{$\tau$}$. 
We follow an approach similar to the 
Kugel-Khomskii formulation.~\cite{Kugel72,Kugel73}
We express the $3d$ electron operators in terms of 
$\vct{$S$}$ and $\vct{$\tau$}$
to arrive at the effective spin and pseudospin Hamiltonian:
$H_{\rm eff} = H_{\rm cry.} + H_{t_{2g}} + H_{e_g}$.      
The first term $H_{\rm cry.}$ denotes the effects of  
$D_{3d}$-crystal field.
This term is obtained from the zeroth-order perturbational processes.
The second term $H_{t_{2g}}$ is
obtained from the second-order perturbational processes whose 
intermediate states contain only $t_{2g}$-orbital degrees
of freedom.
The third term $H_{e_g}$ is obtained 
from the second-order perturbational processes whose 
intermediate states contain $e_g$-orbital degrees
of freedom.  We neglect the relativistic spin-orbit coupling
in the model and discuss this issue later.

\section{Results}

\begin{figure}[tdp]
\hfil
\includegraphics[scale=0.35]{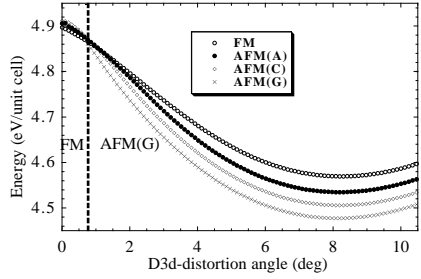}
\hfil
\caption{Energies of each magnetic structures with config. 1 as
functions of the $D_{3d}$-distortion angle. See the text for details.}
\label{eneVMS157}
\end{figure}
\begin{figure}[tdp]
\hfil
\includegraphics[scale=0.35]{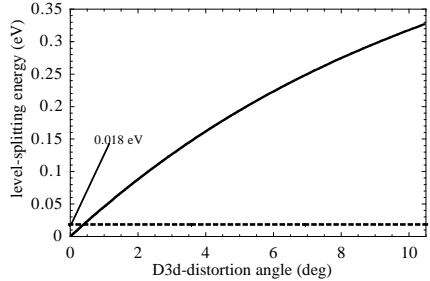}
\hfil
\caption{Level-splitting energy due to the $D_{3d}$-crystal field 
($\Delta_{D_{3d}}$) as a function of the $D_{3d}$-distortion angle.}
\label{lvsplit}
\end{figure}

\begin{fulltable}[tdp]
\caption{Trigonal-axes configurations which lower the AFM(G) spin structure
in energy.}
 \label{tab:configs}
\begin{fulltabular}{ccccc} \hline
      \quad & site 1 & site 2 & site 3 & site 4 \\
      \hline
      \hline
        config.1 & [1,1,1] & [1,1,1] & [1,1,1] & [1,1,1] \\
         $\varphi$ 1  &$\frac{1}{\sqrt{3}}(xy+yz+zx)$
         \quad   &$\frac{1}{\sqrt{3}}(xy+yz+zx)$
         \quad   &$\frac{1}{\sqrt{3}}(xy+yz+zx)$
         \quad   &$\frac{1}{\sqrt{3}}(xy+yz+zx)$            \\
      \hline
        config.2 & [1,-1,1] & [1,-1,1] & [1,-1,1] & [1,-1,1] \\
         $\varphi$ 1  &$\frac{1}{\sqrt{3}}(-xy-yz+zx)$
         \quad   &$\frac{1}{\sqrt{3}}(-xy-yz+zx)$
         \quad   &$\frac{1}{\sqrt{3}}(-xy-yz+zx)$
         \quad   &$\frac{1}{\sqrt{3}}(-xy-yz+zx)$            \\
      \hline
        config.3 & [1,1,1] & [1,1,1] & [1,1,-1] & [1,1,-1] \\
         $\varphi$ 1  &$\frac{1}{\sqrt{3}}(xy+yz+zx)$
         \quad   &$\frac{1}{\sqrt{3}}(xy+yz+zx)$
         \quad   &$\frac{1}{\sqrt{3}}(xy-yz-zx)$
         \quad   &$\frac{1}{\sqrt{3}}(xy-yz-zx)$            \\
      \hline
        config.4 & [1,1,-1] & [1,1,-1] & [1,1,1] & [1,1,1] \\
         $\varphi$ 1  &$\frac{1}{\sqrt{3}}(xy-yz-zx)$
         \quad   &$\frac{1}{\sqrt{3}}(xy-yz-zx)$
         \quad   &$\frac{1}{\sqrt{3}}(xy+yz+zx)$
         \quad   &$\frac{1}{\sqrt{3}}(xy+yz+zx)$            \\
      \hline
        config.5 & [1,-1,1] & [1,-1,1] & [1,-1,-1] & [1,-1,-1] \\
         $\varphi$ 1  &$\frac{1}{\sqrt{3}}(-xy-yz+zx)$
         \quad   &$\frac{1}{\sqrt{3}}(-xy-yz+zx)$
         \quad   &$\frac{1}{\sqrt{3}}(-xy+yz+zx)$
         \quad   &$\frac{1}{\sqrt{3}}(-xy+yz+zx)$    \\      \hline
\end{fulltabular}
\end{fulltable}
There are four trigonal-axes in the ${\rm TiO}_6$ octahedron,
namely, [1,1,1], [1,1,-1], [1,-1,1] and [1,-1,-1]
with respect to the $x$, $y$ and $z$ axes.
At this stage, there are several possibilities for configurations
of the trigonal axes at sites 1, 2, 3, and 4.
The spin-couplings between the neighboring sites depend
on the configuration.
There are five configurations which lower the AFM(G) spin structure
in energy as shown in Table~\ref{tab:configs}.
In this letter, we examine the AFM(G) state with a trigonal-axes 
configuration in which [1,1,1]-axis is chosen at every 
sites (config. 1), and 
we show that a considerably small $D_{3d}$-distortion
does strongly stabilize the AFM(G) state and the experimentally
observed properties are well explained when we assume the distortion.
It is an interesting problem to study which configuration
has the lowest energy, which would require an entirely separate work
and is left for future discussion.
%We calculate the energies of four magnetic structures
%(FM, AFM(A), AFM(C) and AFM(G)) with all possible 
%trigonal-axes-configurations 
%within the mean-field approximation.
%As a result, AFM(G) state with a trigonal-axes configuration
%in which [1,1,1]-axis is chosen at every sites is turned out
%to be the lowest in energy in the region of $\omega > 5^{\circ}$
%where $\omega$ denotes the $D_{3d}$-distortion angle.
%Hereafter, we refer to this configuration of trigonal-axes
%as config. 1.
In the $D_{3d}$ distortion with config. 1, 
the lowest level is $\frac{1}{\sqrt{3}}(xy+yz+zx)$
at every sites.  
In Fig.~\ref{eneVMS157}, we plot the energies of several 
magnetic structures with config.1 as functions 
of $D_{3d}$-distortion angle $\omega$.
In the region of $\omega > 1^{\circ}$, the AFM(G) state is 
strongly stabilized relative to other structures.
This may indeed drive the $D_{3d}$ distortion even in the cost of lattice 
energy. 
Since the lattice elastic-energy is not accounted in Fig.~\ref{eneVMS157},
the real energy minimum including the elastic-energy cost should 
be in the range $0^{\circ}<\omega<8^{\circ}$.
In Fig.~\ref{lvsplit}, we plot the $t_{2g}$-level-splitting energy
due to the $D_{3d}$-crystal field ($\Delta_{D_{3d}}$) as
a function of $\omega$.
In the region of $\omega > 1^{\circ}$, the value of $\Delta_{D_{3d}}$
is sufficiently larger than the coupling constant 
of the relativistic spin-orbit interaction in ${\rm Ti}^{3+}$ ion
(${\xi}_d=0.018$ eV),~\cite{Sugano70} though this distortion angle 
$\omega > 1^{\circ}$ is substantially smaller than the tilting
angle of ${\rm GdFeO}_3$-type distortion.
In fact, it is not surprising if such small $D_{3d}$ distortion
has not been detected so far. 

When the level-splitting energy is sufficiently large
as compared to the characteristic energy of the spin-orbital
exchange interaction, the orbital occupation 
at each site is restricted to $\frac{1}{\sqrt{3}}(xy+yz+zx)$
independently of the magnetic structure.
At this stage, we can estimate the value of the spin-coupling constant
in the $ab$-plane ($J_{\rm Heis}^{ab}$) and that along the $c$-axis 
($J_{\rm Heis}^{c}$) under this assumption.
\begin{figure}[tdp]
\hfil
\includegraphics[scale=0.4]{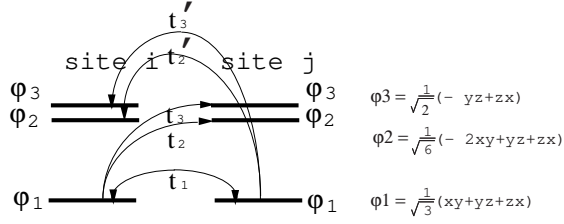}
\hfil
\caption{Characteristic transfers which contribute to the 
spin-exchange interaction. 
%and the expressions of 
%$J_{\rm Heis}^{ab}$ and $J_{\rm Heis}^{c}$ when the orbital
%occupation is restricted to the lowest level.
}
\label{sdproclett}
\end{figure}
In Fig.~\ref{sdproclett},
the characteristic transfers which contribute 
to the spin-exchange interaction are presented.
Using the notation of transfer integrals as shown in Fig.~\ref{sdproclett},
$J_{\rm Heis}^{ab}$ and $J_{\rm Heis}^{c}$ are expressed 
in the strong coupling expansion as,
\begin{equation}
J_{\rm Heis}^{ab}, J_{\rm Heis}^{c} = 
4\frac{t_1^2}{u} 
-2\frac{(t_2^2+t_3^2)j}{(u'+\Delta_{D_{3d}})^2}
-2\frac{({t'}_2^2+{t'}_3^2)j}{(u'+\Delta_{D_{3d}})^2}.         
\end{equation}
\begin{figure}[tdp]
\hfil
\includegraphics[scale=0.4]{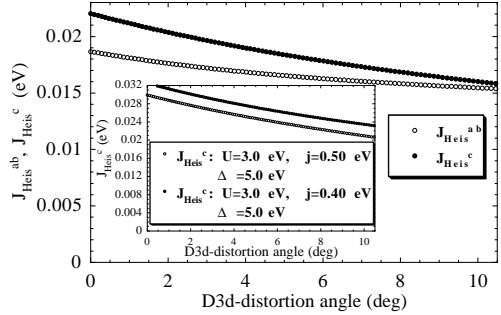}
\hfil
\caption{Spin-exchange interactions estimated under the assumption
that orbital occupation is restricted to the lowest level
($J_{\rm Heis}^{ab}$ and $J_{\rm Heis}^{c}$) are plotted.
Inset shows those calculated using other parameters within the 
error bars of the analyses of photoemission spectra.
See the text for details.}
\label{eJab-Jc}
\end{figure}
In Fig.~\ref{eJab-Jc}, the values of $J_{\rm Heis}^{ab}$
and $J_{\rm Heis}^{c}$ are plotted
as functions of $\omega$.
Both $J_{\rm Heis}^{ab}$ and $J_{\rm Heis}^{c}$ 
take approximately 0.015-0.022 eV, and they are 
in agreement with the value obtained by
the neutron scattering experiment of $\sim 15.5$ meV. 
In ${\rm LaTiO}_3$, the magnetic moment
takes $0.45$ $\mu_{\rm B}$, which is reduced from the spin-only moment.
%of $0.85$ $\mu_{\rm B}$.
Recent optical measurement shows
that ${\rm LaTiO}_3$ has a considerably small optical gap of $\sim 0.1$ eV
in the vicinity of the metal-insulator phase boundary
with strong itinerant character.~\cite{Okimoto95}
Therefore, in this system,
we expect that some amount of orbital and spin fluctuations remain.
The reduction of the magnetic moment may easily be attributed
to these fluctuations.
We note that in this context, the calculated
$J_{\rm Heis}^{ab}$ and $J_{\rm Heis}^{c}$ in the localized limit
seem rather small at first sight.
To elucidate this issue, we have examined 
how the values of $J_{\rm Heis}^{ab}$ and $J_{\rm Heis}^{c}$
change within the error bars of the parameters estimated from the analyses
of photoemission spectra.
In ${\rm LaTiO}_3$ with strong
itinerant character, we can expect that the Coulomb and exchange
interactions are rather small.
We calculate the values in the case of $U = 3.0$ eV, 
$j = 0.50$ eV and $\Delta = 5.0$ eV and in the case of $U = 3.0$ eV,
$j = 0.40$ eV and $\Delta = 5.0$ eV.
The values take $\sim$ 0.03 eV, 
which are larger than the previous estimate (see Inset of Fig.~\ref{eJab-Jc}).
Considering these variations and uncertainties, we conclude
that the calculated $J_{\rm Heis}^{ab}$ and $J_{\rm Heis}^{c}$
in the localized limit are in qualitative agreement with the 
experimentally obtained $J$ in the presence of itinerant
charge fluctuations. 
 
Here, a question arises: why has the orbital ordering caused by
the occupation of the lowest level in the $D_{3d}$ distortion
not been detected experimentally so far?
Detection of the orbital ordering by the resonant x-ray scattering
is based on the splitting of the Ti $4p$ levels induced by the 
orbital ordering of the Ti $3d$ states.
When the Ti $3d$ orbitals are ordered in the $D_{3d}$ distortion,
the Coulomb interaction between Ti $4p$ and occupied $3d$ orbitals,
$U_{4p-3d}$ and the interaction
between Ti $4p$ and ligand O $2p$ orbitals, $U_{4p-2p}$
make opposite contributions.
Among the threefold Ti $4p$ orbitals, the orbital directed along the trigonal
axis has higher energy due to $U_{4p-3d}$.
On the contrary, along this axis, $U_{4p-2p}$ is weakened
so that the crystal field works to lower the Ti $4p$ orbital.
In addition, the weakend hybridizations between the Ti $4p$ and O $2p$
orbitals might lower the Ti $4p$ orbital.
This cancellation may lead to the absence of detectable orbital ordering 
in the resonant x-ray scattering 
in ${\rm LaTiO}_3$.~\cite{Keimer00}
The orbital ordering accompanied by the $D_{3d}$ distortion
would be more easily detected in the AFM(G) compounds with relatively large
optical gap and small charge fluctuations: ${\rm PrTiO}_3$ and ${\rm NdTiO}_3$.

In summary, in order to interpret the experimentally observed 
properties of AFM(G) state in ${\rm LaTiO}_3$,
we have examined the effects of possible $D_{3d}$-crystal field
as a candidate for their origin.
In the presence of the $D_{3d}$-crystal field, 
threefold cubic-$t_{2g}$-levels
are split into non-degenerate $a_{1g}$- and 
twofold-degenerate $e_g$-levels.
Since the lowest level has no degeneracy, this distortion 
strongly lowers the energy of the electron system in $t_{2g}^1$ configuration.
%So that, though the $d$-type JT distortion is realized in the 
%large ${\rm GdFeO}_3$-type distortion or in ${\rm YTiO}_3$
%due to , we can expect that the $D_{3d}$ distortion is 
%realized in ${\rm LaTiO}_3$
%with small ${\rm GdFeO}_3$-type distortion.
Moreover, the calculated $D_{3d}$-splitting energy $\Delta_{D_{3d}}$ suggests
that the $D_{3d}$-splitting dominates over the relativistic
spin-orbit interaction in the region of 
$D_{3d}$-distortion angle $\omega > 1^{\circ}$.
The calculated spin-exchange constant is in agreement with that
obtained experimentally.
%On the basis of energy calculations of 
%magnetic structures with several trigonal-axes configurations,
%the AFM(G) state with config. 1 in which [1, 1, 1]-axis is chosen
%as the trigonal axis at sites 1, 2, 3 and 4 is proved to
%be the lowest in energy in the region of $\omega > 5^{\circ}$.
%In this AFM(G) state, all sites are occupied by  
%$\frac{1}{\sqrt{3}}(xy+yz+zx)$. 
%This orbital occupancy can be attributed to the fact that 
%no evidence of orbital ordering has not been detected 
%by the resonant X-ray scattering.
Finally, our proposal for the small $D_{3d}$ distortion 
should in principle be observed by the displacement of O ions if
the experimental resolution is sufficient.

M. M. thanks H. Asakawa and T. Mizokawa for useful comments. 
This work is supported by ``Research for the Future Program''
(JSPS-RFTF97P01103) from the Japan Society for the Promotion of Science.
%%%%%%

\end{document}